\documentclass[aps,prd,preprint,superscriptaddress,tightenlines,nofootinbib,showpacs]{revtex4}
\usepackage{graphicx}
\usepackage{dcolumn}
\usepackage{bm}

\begin{document}
\preprint{CLNS 05/1940}       
\preprint{CLEO 05-28}         

\title{Radiative Decays of the $\Upsilon(1S)$ to
       $\gamma\pi^0\pi^0$, $\gamma\eta\eta$ and $\gamma\pi^0\eta$}

\author{D.~Besson}
\affiliation{University of Kansas, Lawrence, Kansas 66045}
\author{T.~K.~Pedlar}
\affiliation{Luther College, Decorah, Iowa 52101}
\author{D.~Cronin-Hennessy}
\author{K.~Y.~Gao}
\author{D.~T.~Gong}
\author{J.~Hietala}
\author{Y.~Kubota}
\author{T.~Klein}
\author{B.~W.~Lang}
\author{R.~Poling}
\author{A.~W.~Scott}
\author{A.~Smith}
\affiliation{University of Minnesota, Minneapolis, Minnesota 55455}
% ...
\collaboration{CLEO Collaboration} 
\noaffiliation

\date{\today}

\begin{abstract}
We report on a study of exclusive radiative decays of the $\Upsilon(1S)$ 
resonance into the final states $\gamma\pi^0\pi^0$, $\gamma\eta\eta$ and 
$\gamma\pi^0\eta$, using 1.13 fb$^{-1}$ of $\mathrm{e^+ e^-}$ annihilation
data collected at $\sqrt{s}= 9.46\,$GeV with the CLEO III detector operating at
the Cornell Electron Storage Ring. In the channel $\gamma\pi^0\pi^0$, we measure
the branching ratio for the decay mode 
$\Upsilon(1S) \rightarrow \gamma f_2(1270)$ to be 
$(10.5 ~\pm 1.6~(\mathrm{stat}) ~~^{+ 1.9}_{-1.8}~(\mathrm{syst})) \times 10^{-5}$. We place 
upper limits on the product branching ratios for the isoscalar resonances 
$f_0(1500)$ and $f_0(1710)$ for the $\pi^0\pi^0$ and $\eta\eta$ decay channels.
We also set an upper limit on the 
$\Upsilon(1S)$ radiative decay into $\pi^0\eta$.
\end{abstract}

\maketitle
Radiative decays of quarkonia, where one of the three gluons arising from the quark-antiquark
annihilation is replaced by a photon leaving two gluons to form bound states, are
thought to be a glue-rich environment that may lead to the production of 
glueballs and gluonic-mesonic states rather than ordinary mesons \cite{bib:1,bib:1a}. 
Lattice gauge
theory calculations \cite{bib:1b,bib:2} predict that the lightest glueball
should have $J^{PC} = 0^{++}$ and that its mass should be in the range of 1.45
to 1.75 GeV/$c^2$, with decay into two pseudo-scalars 
($J^{PC} = 0^{-+}$) expected to dominate.
Unfortunately, the identification of a scalar glueball among the many
established scalar resonances is difficult, as they
have the same quantum numbers and similar decay modes 
and may mix.
The triplet of $f_0$ states
are likely candidates for the superposition of quark states and a
scalar glueball state. 
Many of the lattice QCD models predict the decay ratios
(e.g.,~$\eta\eta / \pi\pi$, $\eta\eta / K\bar{K}$) for a glueball and for
scalar resonances \cite{bib:4}, and this is a possible
tool to distinguish among them.

Most of the information on radiative decays of quarkonia has centered
on $J/\psi$ decays \cite{bib:9,bib:10,bib:11,bib:12,bib:13}, leading to
a list of two-body decay branching ratios. The establishment of a corresponding
list for $\Upsilon(1S)$ decays is desirable and would not only
deepen our understanding of $c\bar{c}$ and $b\bar{b}$ quarkonia, but could also
contribute to the identification of a scalar glueball state or shed new
light on its mixing with ordinary nearby meson states.

Recently, radiative decays into two charged particles have been studied
by the CLEO~III collaboration \cite{bib:18}.
The analysis
included a measurement of the decay rate into $f_2(1270)$, 
a confirmation of its spin, and a measurement
of its helicity distribution. In this analysis, we use the same CLEO~III
$\Upsilon(1S)$ data sample to perform a complementary study of 
all-neutral decays.
Although these final states are subject to poorer resolution and
efficiency than those with charged particles, they have the advantage
of having no
background from QED final states such as $\gamma\rho$. Furthermore, they 
allow the search for states decaying into $\eta\eta$ and 
$\pi^0\eta$. 
Resonant production in the latter mode would be a signature of unexpected physics.

The analysis presented here uses data collected by the CLEO~III detector
configuration \cite{bib:19,bib:22}
at the Cornell Electron Storage Ring (CESR).
The vital component for this analysis is the  CsI(Tl) calorimeter, which 
has a resolution of 1.5\%(2.2\%) for 1 GeV(5 GeV) photons, typical 
of the photons studied here.
We search for radiative $\Upsilon(1S)$ decays in the modes
$\Upsilon(1S) \rightarrow \gamma \pi^0\pi^0$, $\gamma\eta\eta$ and
$\gamma\pi^0\eta$. The $\Upsilon(1S)$ data ($\mathrm{E_{cm}}$ = 9.46 GeV) sample
consists
of an integrated luminosity of 1.13 fb$^{-1}$, corresponding to
$(21.2 \pm 0.2(\mathrm{syst})) \times 10^{6}$ $\Upsilon(1S)$ decays
\cite{bib:26b}.

Candidate events for the individual final states 
($\gamma \pi^0\pi^0$, $\gamma \eta\eta$ and $\gamma \pi^0\eta$)
are selected in a similar fashion, using the following basic selection criteria.
An event must have no charged tracks and exactly one electromagnetic
shower in the barrel ($|\cos\theta| <$ 0.75, where $\theta$ represents the
polar angle)
or the endcap region
(0.82 $< |\cos\theta| <$ 0.93) of the calorimeter with an energy exceeding 
4 GeV, together with at least four other photons in the event. 
All combinations of two photons (excluding the photon that has $E>4\ $GeV) in the
event are then combined to form $\pi^0$ and $\eta$ candidates.
To be selected, an event must have two pairs of photons satisfying the 
requirement

\[ \sqrt{P_1^2(\pi_1^0/\eta_1) +  P_2^2(\pi_2^0/\eta_2)} < 5, \]

\noindent
with $P_1$ and $P_2$ being the pulls, defined as:

\[ P(\pi^0/\eta) = \left [ m_{\gamma\gamma} - m(\pi^0/\eta) \right ] / 
                         \sigma_{\gamma\gamma}, \]

\noindent
where $m_{\gamma\gamma}$ is the $\gamma\gamma$ invariant mass, $m(\pi^0/\eta)$ 
is the known
$\pi^0$ or $\eta$ mass, \cite{bib:28} 
and $\sigma_{\gamma\gamma}$ is the  
$\gamma\gamma$ mass resolution, with typical values of 5 - 7 MeV/$c^2$.
The $\pi^0$ and $\eta$ candidates are then kinematically constrained 
to their masses, $m(\pi^0)$ and $m(\eta)$.

To study the event-selection criteria and measure their efficiencies, we use a 
Monte Carlo
simulation consisting of an event generator \cite{bib:27a} and a GEANT-based
\cite{bib:27} detector-response simulation. For each final state,
$\Upsilon(1S) \rightarrow \gamma X$, events are generated with $X =
f_2(1270)$, $f_0(1500)$ and $f_0(1710)$, using a Breit-Wigner 
line-shape
and the PDG mass and width
\cite{bib:28}. We do not search for the $f_0(1370)$ as it overlaps completely
in mass with the $f_2(1270)$ due to its large instrinsic width.

\begin{figure}[htb]
\includegraphics*[width=3.4in]{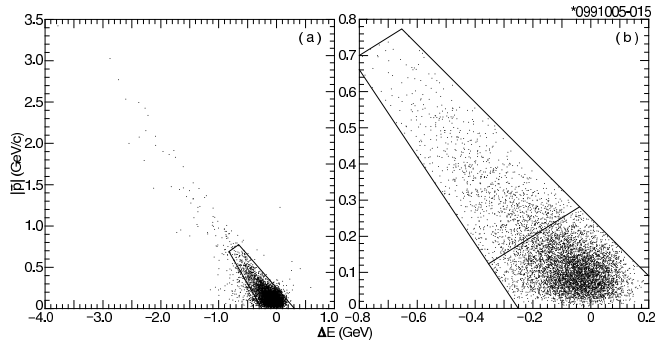}
\caption{Illustration of the chosen 4-momentum distribution cuts on a larger (a) and
smaller (b) scale, using Monte Carlo events for the decay channel
$\Upsilon(1S) \rightarrow \gamma  f_2(1270) \rightarrow \gamma \pi^0\pi^0$.
The slanted line in the lower right part of (b) divides the selected events
roughly into two categories: the lower area where the 4-momentum is conserved
for the entire event; the upper area where the 4-momentum is conserved for the
intermediate resonance but not for the entire event. The remaining events
outside the selected area in (a) have an energy loss from more than one photon
and, hence, are excluded from the selection.}
\label{fig:1}
\end{figure}

A 4-momentum cut 
and an
asymmetry cut are then used to further select candidate events. 
For the $\gamma\pi^0\pi^0$ final-state selection, the allowed region for the
4-momentum is bounded by the following three conditions:
$ |\vec{p}|=-0.30-1.20~\Delta E,  
|\vec{p}|=0.25 - 0.80\Delta E$, and
$|\vec{p}|   = 1.10 + 0.50~\Delta E$,
where $\Delta E$ is the difference between the reconstructed event energy
and the center-of-mass energy ($\mathrm{E_{cm}}$)
in GeV and $|\vec{p}|$ is the magnitude of the reconstructed total event
momentum in GeV/$c$.
These cuts include the $\Delta E - |\vec{p}|$ area where the 4-momentum
is conserved for the
entire event and increase the efficiency by, in addition, including
the region where
the single, recoiling photon is reconstructed with too low an energy. 
For the latter, the
4-momentum is not conserved for the entire event but only for the intermediate
resonance $X$ in the decay chain
$\Upsilon(1S) \rightarrow \gamma X \rightarrow \gamma \pi^0\pi^0$.
These cuts are illustrated in Figure~\ref{fig:1}.
We define a 4-momentum allowed region 
for the $\gamma\eta\eta$ final state selection in a similar manner.

A source of background originates from combining a wrong pair of photons
to form
a $\pi^0$ or $\eta$ candidate. 
Real $\pi^0$ and $\eta$ mesons decay isotropically and their angular
distributions are flat.
However, the $\pi^0$ and $\eta$ candidates which originate from a wrong photon 
combination 
do not have a flat
distribution in this variable and can largely be removed by a cut which uses
the polar $\Delta\theta$ and
azimuthal $\Delta\varphi$ angle differences between the
two photons from a decay candidate. 
For the $\gamma\pi^0\pi^0$ final state the asymmetry requirement is
$\sqrt{\Delta\theta^2 + \Delta\varphi^2} < 40^{\circ}$, while for the 
$\gamma\eta\eta$
final state it is
$\sqrt{\Delta\theta^2 + \Delta\varphi^2} < 60^{\circ}$. 

Comparison of the invariant $\pi^0\pi^0$ and $\eta\eta$ 
mass spectra from the Monte Carlo simulation 
reveals significant differences in the mass and width values from the ones 
used at the
generator level. These differences are parametrized in the form of 
Gaussian resolution functions
off-set from zero.
The mass shift is an artifact of the shower reconstruction
in the calorimeter of such fast $\pi^0$ and $\eta$ mesons,
when the showers tend to overlap.
For example, the resolution function for the decay $f_2(1270)\to\pi^+\pi^-$
is a Gaussian function with $\sigma=\ $ 27 ${\rm MeV/c^2}$ and an offset of 
$-20\ {\rm MeV/c^2}$.

We determine the selection efficiency for each of the resonances individually. The
event selection efficiencies 
are summarized in
Table~1. The uncertainties shown are statistical only.

\begin{table}[h]
\begin{tabular}{@{}c|c|c}
\hline
\hline
~Resonance~ & \multicolumn{2}{c}{~Reconstruction Efficiency in \%~} \\
          & $\gamma\pi^0\pi^0$ & $\gamma\eta\eta$ \\
\hline
$f_2(1270)$ & ~~~16.4 $\pm$ 0.2~~~ & 10.2 $\pm$ 0.2 \\
$f_0(1500)$ & 20.4 $\pm$ 0.3 &  9.1 $\pm$ 0.2 \\
$f_0(1710)$ & 20.6 $\pm$ 0.3 &  8.6 $\pm$ 0.2 \\
\hline
\hline
\end{tabular}
\caption
{Reconstruction efficiencies for various inter\-mediate resonances in the
$\gamma\pi^0\pi^0$ and $\gamma\eta\eta$ final states.
}
\label{tab:2}
\end{table}

The major background contribution in our signal region originates
from non-resonant
processes. CLEO's
sample of data collected in the continuum below the $\Upsilon(1S)$
$(192 {\rm pb^{-1}})$ is too small to perform a continuum
subtraction.
Hence, we parametrize the background using a threshold 
function of the 
form
\begin{displaymath}
F(x) = N \cdot (x - T) \cdot e^{c_1 (x - T) ~+ ~c_2 (x - T)^2},
\end{displaymath}
where $x$ is the $\pi^0\pi^0$ invariant mass, $N$ is a scale factor,
$T$ is the mass threshold, 
and $c_1$, $c_2$ are free parameters.
This functional form is a good fit to the spectrum obtained 
from a large Monte Carlo data sample of continuum events, and also
to continuum events taken at energies near the $\Upsilon(4S)$.

Figure~2 shows the final $\pi^0\pi^0$ and $\eta\eta$ invariant mass 
spectra. 
The $\pi^0\pi^0$ invariant mass distribution is dominated by the
isoscalar resonance $f_2(1270)$. The $\eta\eta$ invariant mass distribution,
Figure~2(b), has only two events, which is too few
to show any resonant structure.

The Monte Carlo signal events for the processes
$\Upsilon(1S) \rightarrow \gamma ~X \rightarrow \gamma \pi^0\pi^0/\eta\eta$
are produced with a decay angle distribution which is characteristic of the
spin of the final state
($i.e.,\ J = 0$ for $f_0(1500)$ and $J = 2$ for $f_2(1270)$). However, the
generation does not take into account the correct helicity distribution 
for the $f_2(1270)$ since
this
distribution depends on the specific decay channel and
can only
be determined from the data itself.
The method to obtain the correct helicity-angle distributions
is described in detail in \cite{bib:18} and results in a helicity
correction factor
which takes into account decay-dependent 
efficiency corrections and the helicity 
substructure for the final state resonance. For this analysis,
we use the helicity substructure, which is independent 
of the charge of the pions,  determined in \cite{bib:18},
as this is more precise than the one we can determine
using the decay into $\pi^0\pi^0$.
We obtain
a correction factor for the $f_2(1270)$
of 0.66 $\pm$ 0.04,
where the uncertainty is
statistical only.
This factor multiplies the efficiency stated in Table \ref{tab:2}.

To determine the branching ratio for
$\Upsilon(1S) \rightarrow \gamma f_2(1270)$,
we fit the invariant $\pi^0\pi^0$
mass distribution with a spin-2
Breit-Wigner line-shape of fixed mass and width, convolved
with the resolution function derived from Monte Carlo studies
as previously described,
together with the threshold function.
Integrating the Breit-Wigner line-shape fit from 0.28 to 3.0 GeV/$c^2$
gives
67.9 $\pm$ 10.2 events for the $f_2(1270)$.

\begin{figure}[htb]
\includegraphics*[width=3.4in]{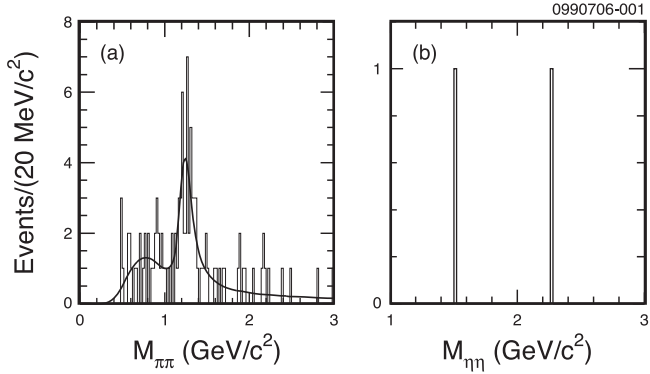}
\caption{The (a) $\pi^0\pi^0$
invariant mass distribution, and (b) $\eta\eta$ invariant mass distribution, 
from the $\Upsilon(1S)$ data sample. The line shows the fit described in the 
text.
}
\label{fig:7}
\end{figure}

With the results from this line-shape fit, the efficiency from 
Table \ref{tab:2}, and the helicity-correction factor, we
determine the product branching ratio for the $f_2(1270)$ to be:

\medskip
\centerline{${\cal B}(\Upsilon(1S) \rightarrow \gamma f_2(1270)) \cdot
{\cal B}(f_2(1270) \rightarrow \pi^0\pi^0) = (3.0 \pm 0.5) \times 10^{-5}$,}
\par
\medskip
\noindent
where the error is statistical only.

We determine a systematic uncertainty on this branching ratio of $\pm$17\%. 
The largest
contribution to this error originates from uncertainties in the line-shape fit
and the threshold function used for the background parametrization. Other 
contributions include systematic uncertainties in the $\pi^0$
reconstruction and in the 4-momentum cut.
Taking into account the isoscalar nature of the $f_2(1270)$ and the
branching ratio of
${\cal B}(f_2(1270) \rightarrow \pi\pi) = 0.847^{+ 0.025}_{-0.012}$
\cite{bib:28}, we determine an overall $\Upsilon(1S)$ radiative decay 
branching ratio to 
$f_2(1270)$ of:

\bigskip

\centerline{${\cal B}(\Upsilon(1S) \rightarrow \gamma f_2(1270)) =
(10.5 ~\pm 1.6~(\mathrm{stat}) ~~^{+ 1.9}_{-1.8}~(\mathrm{syst})) \times 10^{-5}$.}
\par
\medskip

To set upper limits on the branching ratios for other likely
resonances in the $\gamma\pi^0\pi^0$
final state, we
include an additional spin-dependent Breit-Wigner line-shape in the $f_2(1270)$
branching ratio fit, with
a line-shape determined from our Monte Carlo studies. We fix
the area of the additional Breit-Wigner and then repeat the fit
using different values for the number of events.  We then
plot the number of events versus their likelihood from the fit, numerically
integrate the area under the curve and determine the
number of events where
90\% of the physically allowed area is covered. 
This number represents the upper limit
at the 90\% confidence level (C.L.), which
we find 
to be 6.9 events
for the 
$f_0(1500)$
and to be 6.6 events for the $f_0(1710)$.
Using the
branching ratio
${\cal B}(f_0(1500) \rightarrow \pi\pi) = 0.349 \pm 0.023$ \cite{bib:28},
and incorporating the systematic uncertainties ($\approx$6\%)  
in the efficiencies by
smearing the probability density function,
we determine the 90\% C.L. upper limit branching ratio
for the $f_0(1500)$ to be
\begin{displaymath}
{\cal B}(\Upsilon(1S) \rightarrow \gamma f_0(1500)) < 1.5 \times 10^{-5},
\end{displaymath}
and the product branching ratio for the $f_0(1710)$ to be
\begin{displaymath}
{\cal B}(\Upsilon(1S) \rightarrow \gamma f_0(1710)) \cdot
{\cal B}(f_0(1710) \rightarrow \pi^0\pi^0) < 1.4 \times 10^{-6}.
\end{displaymath}

As we see no evidence of any resonant 
structure in the $\eta\eta$ invariant mass distribution
we measure upper limit branching ratios for the 
$f_0(1500)$ and $f_0(1710)$. For this determination we use the
simple method of event counting. The final invariant mass plot has
negligible background and, hence, we assume both events are from
the $\Upsilon(1S) \rightarrow \gamma \eta\eta$ final state. Therefore,
the number
of events
follows a Poisson distribution.
For the $f_0(1500)$ we find 1 event in the mass interval of 1 full-width around
its mass and 0 events for the $f_0(1710)$,
which translates into 90\% C.L. upper limits
of 3.9 and 2.3 events, respectively. 

The systematic uncertainty for the $f_0(1500)$ and the
$f_0(1710)$ is $\approx$ 30\%.
The largest contributions to these uncertainties originate from the
4-momentum requirement, and the $\eta$ asymmetry cut, 
which are on the order of
20\%. Combining the statistical and systematical uncertainties, we determine
the 90\% C.L. upper limit on the product branching ratio for the $f_0(1500)$ to be:
\begin{displaymath}
{\cal B}(\Upsilon(1S) \rightarrow \gamma f_0(1500)) \cdot
{\cal B}(f_0(1500) \rightarrow \eta\eta) < 3.0 \times 10^{-6},
\end{displaymath}
and for the $f_0(1710)$ to be:
\begin{displaymath}
{\cal B}(\Upsilon(1S) \rightarrow \gamma f_0(1710)) \cdot
{\cal B}(f_0(1710) \rightarrow \eta\eta) < 1.8 \times 10^{-6}.
\end{displaymath}

In the decay $\Upsilon(1S) \rightarrow \gamma X$, 
if we assume that the $\gamma$ is produced directly and is not the product
of an intermediate virtual particle, the resonance $X$ must be
an iso-scalar. In this case, if $X$ is conventional meson state,
it can only decay into a pair of pseudo-scalars ($J^P = 0^-$) each with $I$ = 0 
($e.g.$, $\eta\eta$), or
$I$ = 1, $e.g.,\ \pi\pi$. 
Observation of a resonance in $\pi^0\eta$ could therefore be an indication
that the photon in this case is the result of enhanced production 
via an intermediate hadron, or alternatively the result of an 
unexpectedly large I=0 component of the $\pi^0\eta$ final state.

Following the same analysis chain as detailed above and using 
the same $\vec{p},\Delta E$ region 
as for the $\pi^0\pi^0$ case, we
find no events in our signal region for this decay.
Hence, we determine an upper limit for the branching ratio
$\Upsilon(1S) \rightarrow \gamma \pi^0 \eta$.

We use the same method as for the upper
limit determination in the $\Upsilon(1S) \rightarrow \gamma \eta \eta$
final state. To measure the
reconstruction efficiency for
any exotic-state mass, we generate Monte Carlo events of the type
$\Upsilon(1S) \rightarrow \gamma \pi^0 \eta$ with a flat $\pi^0 \eta$ 
invariant mass
distribution between 0.7 and 3 GeV/$c^2$, and use the lowest efficiency
found in the entire mass distribution of (4.8 $\pm$ 0.5)\%; The 
efficiency is relatively
flat over the mass interval of interest.

Having no events in the data over the mass range of 0.7 to 3.0 GeV/$c^2$
corresponds to a
90\% C.L. upper limit of 2.3 events.
Combining this with a systematic error of $^{+24}_{-14}\%$, 
due to the same sources of uncertainty
as with the previous 2 analyses, we determine the 90\% C.L. 
upper limit for the branching ratio 
to be:
\begin{displaymath}
{\cal B}(\Upsilon(1S) \rightarrow \gamma \pi^0\eta) < 2.4 \times 10^{-6}.
\end{displaymath}

In summary,
we have analyzed 1.13 fb$^{-1}$ of data from the CLEO~III detector at the
$\Upsilon(1S)$ for resonances in the radiative decay channels 
$\Upsilon(1S) \rightarrow \gamma \pi^0 \pi^0$, $\gamma \eta \eta$ and 
$\gamma \pi^0 \eta$.

In the decay channel $\gamma\pi^0\pi^0$, we measure a branching ratio value for
the isoscalar resonance $f_2(1270)$ of
${\cal B}(\Upsilon(1S) \rightarrow \gamma f_2(1270)) =
(10.5 ~\pm 1.6 ~~^{+ 1.9}_{-1.8}) \times 10^{-5}$.
This is in excellent agreement with
the same branching ratio obtained from the charged final state 
$\gamma\pi^-\pi^+$, using the
same CLEO~III data set:
${\cal B}(\Upsilon(1S) \rightarrow \gamma f_2(1270)) =
  (10.2 ~\pm 0.8 ~\pm 0.7) \times 10^{-5}$ \cite{bib:18}. 
It also agrees within the uncertainties with the earlier
CLEO~II result of $(8.1 ~\pm 2.3 ~\pm 2.7) \times 10^{-5}$,
based on the decay channel $\gamma\pi^-\pi^+$ \cite{bib:17};
this earlier measurement had no correction for the helicity
distribution, and the large systematic uncertainty reflected this fact.

In addition, we determine 90\% C.L. upper limits
for the isoscalar resonances $f_0(1500)$ and $f_0(1710)$ decaying into
$\pi\pi$, as well as a 90\% C.L. upper limit for the decay
$\Upsilon(1S) \rightarrow \gamma f_0(1500)$.
Based on the scalar-glueball mixing matrix from \cite{bib:4},
QCD factorization model calculations in \cite{bib:1a}
predict branching ratios for the $f_0(1500)$ and
$f_0(1710)$ to be
${\cal B}(\Upsilon(1S) \rightarrow \gamma f_0(1500)) \approx 
42 - 84 \times 10^{-5}$
and 
${\cal B}(\Upsilon(1S) \rightarrow \gamma f_0(1710)) \cdot
{\cal B}(f_0(1710) \rightarrow \pi^0\pi^0) \approx 6 - 12 \times 10^{-6}$.
Our measurements of
${\cal B}(\Upsilon(1S) \rightarrow \gamma f_0(1500)) < 1.5 \times 10^{-5}$ and
${\cal B}(\Upsilon(1S) \rightarrow \gamma f_0(1710)) \cdot
{\cal B}(f_0(1710) \rightarrow \pi^0\pi^0) < 1.4 \times 10^{-6}$
are much
smaller than these predictions.

In the $\gamma\eta\eta$ decay channel, no resonant structures are observed.
Therefore, we determine a 90\% C.L.
upper limit on the branching ratios for the isoscalar
resonances $f_0(1500)$ and $f_0(1710)$ decaying into
$\eta\eta$ as
${\cal B}(\Upsilon(1S) \rightarrow \gamma f_0(1500)) \cdot
{\cal B}(f_0(1500) \rightarrow \eta\eta) < 3.0 \times 10^{-6}$ and
${\cal B}(\Upsilon(1S) \rightarrow \gamma f_0(1710)) \cdot
{\cal B}(f_0(1710) \rightarrow \eta\eta) < 1.8 \times 10^{-6}$.

The search for states in the $\gamma\pi^0\eta$ decay channel does not
show any evidence of a signal.
We determine a 90\% C.L. upper limit on the branching ratio for the
decay $\Upsilon(1S) \rightarrow \gamma \pi^0 \eta$ for any intermediate state
with a mass between 0.7 and 3.0 GeV/$c^2$ to be
${\cal B}(\Upsilon(1S) \rightarrow \gamma \pi^0\eta) < 2.4 \times 10^{-6}$.

\par
\bigskip
We gratefully acknowledge the effort of the CESR staff
in providing us with excellent luminosity and running conditions.
A.~Ryd thanks the A.P.~Sloan Foundation.
This work was supported by the National Science Foundation
and the U.S. Department of Energy.


\begin{thebibliography}{99}
\bibitem{bib:1} J.~D.~Bjorken, Proceedings of Summer Institute of Particle 
Physics, SLAC-PUB 224, 219 (1980).

\bibitem{bib:1a}    X.-G. He, H.-Y. Jin and J.P. Ma,
                    Phys. Rev. {\bf D66} (2002) 074015.
\bibitem{bib:1b}     C.J. Morningstar {\it et al.},
                    Phys. Rev. {\bf D60} (1999) 034509.
\bibitem{bib:2}     C.J. Morningstar {\it et al.},
                    AIP Conf. Proc. {\bf 688} (2004) 220.
\bibitem{bib:4}     F.E. Close and A. Kirk,
                    Eur. Phys. J. {\bf C21} (2001) 531.
\bibitem{bib:9}     J.E. Augustin {\it et al.} (DM2 Collaboration),
                    Z. Phy. {\bf C36} (1987) 369.
\bibitem{bib:10}    R.M. Baltrusaitis {\it et al.} (Mark~III Collaboration),
                    Phys. Rev. {\bf D35} (1987) 2077.
\bibitem{bib:11}    C. Edwards {\it et al.} (Crystal Ball Collaboration),
                    Phys. Rev. Lett. {\bf 48} (1982) 458.
\bibitem{bib:12}    J.Z. Bai {\it et al.} (BES Collaboration),
                    Phys. Rev. Lett. {\bf 81} (1998) 1179.
\bibitem{bib:13}    X.Y. Shen {\it et al.} (BES Collaboration),
                    Proceedings of Physics in Collision 2002,
                    eConf {\bf C020620:THAT07} (2002).
\bibitem{bib:18}    S.B. Athar {\it et al.} (CLEO Collaboration),
                    Phys. Rev. {\bf D73} (2006) 032001.
\bibitem{bib:19}    Y. Kubota {\it et al.} (CLEO Collaboration),
                    Nucl. Inst. Meth. {\bf A320} (1992) 66.
\bibitem{bib:22}    G. Viehhauser {\it et al.},
                    Nucl. Inst. Meth. {\bf A462} (2001) 146. 
\bibitem{bib:26b}   R.A. Briere {\it et al.} (CLEO Collaboration),
                    Phys. Rev. {\bf D70} (2004) 072001.
\bibitem{bib:28}    Y.-M.Yao {\it et al.}, Review of Particle Physics,
                    J. Phys. {\bf G33} (2006) 1.
\bibitem{bib:27a}   {\it QQ - The CLEO Event Generator}, \\
                    http://www.lns.cornell.edu/public/CLEO/soft/QQ
                    (unpublished).
\bibitem{bib:27}    R. Brun et al., GEANT 3.21,
                    CERN Program Library Long Writeup W5013 (1993).
\bibitem{bib:17}    A. Anastassov {\it et al.} (CLEO Collaboration),
                    Phys. Rev. Lett. {\bf 82} (1999) 286.

\end{thebibliography}
\end{document}